\documentclass[12pt]{article}
\usepackage[left=2.8cm,right=2.8cm,top=2.8cm,bottom=2.8cm]{geometry}
\usepackage[colorlinks=true,linkcolor=red,citecolor=blue,urlcolor=green,filecolor=black]{hyperref}

\usepackage{tensor}
\usepackage{amstext,amssymb,amsfonts,amsmath,amsthm}
\usepackage{textcomp}
\usepackage{mathrsfs}
\usepackage{multirow}
\usepackage{array}
\usepackage{graphicx}
\usepackage{calc}
\usepackage{color}
\usepackage[nosort]{cite}
\usepackage[normalem]{ulem}
\renewcommand{\theequation}{\thesection.\arabic{equation}} \csname
@addtoreset\endcsname{equation}{section}

\def\be{\begin{equation}}
\def\ee{\end{equation}}
\def\bea{\begin{eqnarray}}
\def\eea{\end{eqnarray}}
\def\ba{\begin{array}}
\def\ea{\end{array}}
\def\bwt{\begin{widetext}}
\def\ewt{\end{widetext}}
\def\nn{\nonumber\\}

\newcommand{\f}[2]{\frac{#1}{#2}}
\newcolumntype{C}[1]{>{\centering}m{#1}}

\begin{document}

%\begin{flushright}
%\begin{tabular}{c}
%Preprint number
%\end{tabular}
%\end{flushright}

\vspace{30pt}

\begin{center}

%%%%%%%%%%%%%%%%%%%%%%%%%%%%%%%%%%%%%%%%%%%%%%%%%%%%%%%%%%%%%%%%%%%%
{\Large 
Flat $\mathfrak{so}(p,q)$-Connections for Manifolds of Non-Euclidean Signature}
%%%%%%%%%%%%%%%%%%%%%%%%%%%%%%%%%%%%%%%%%%%%%%%%%%%%%%%%%%%%%%%%%%%%

\vspace{25pt}
{ Arash Ranjbar${}^{\, a, b}$ and Jorge Zanelli${}^{\,  b}$}

\vspace{10pt}
{${}^a$\sl\small
Universit\'e Libre de Bruxelles and International Solvay Institutes, ULB-Campus Plaine CP231, B-1050 Brussels, Belgium\\
\vspace{10pt}
${}^b$\sl \small Centro de Estudios Cient\'{i}ficos (CECs), Arturo Prat 514, Valdivia, Chile}
%%%%%%%%%%%%%%%%%%%%%%%%%%%%%%%%%%%%%%%%%%%%%%%
\vspace{40pt}

{\sc\large Abstract} 
\end{center}
\noindent

The well-known fact that $S^1$, $S^3$ and $S^7$ are parallelizable manifolds admitting flat connections is revisited. The role of torsion in the construction of those flat connections is made explicit, and the possibilities allowed by different metric signatures are examined. A necessary condition for parallelizability in an open region is that the torsion tensor must be covariantly constant. This property can be used to obtain a relation between a torsion-free and flat connections. Our treatment covers Riemannian and pseudo-Riemannian (non-Euclidean signature) hyperbolic manifolds of dimensions three and seven. Apart from the spherical cases mentioned above, the explicit flat $\mathfrak{so}(p,q)$ connections with $p+q=3,7$ and $p-q=1$ are constructed for the coset manifolds $SO(p,q+1)/SO(p,q)$.

%%%%%%%%%%%%%%%%%%%%%%%%%%%%%%%%%%%%%%%%%%%%%%%%%
\newpage
%%%%%%%%%%%%%%%%%%%%%%%%%%%%%%%%%%%%%%%%%%%%%%%%%
\tableofcontents
%%%%%%%%%%%%%%%%%%%%%%%%%%%%%%%%%%%%%%%%%%%%%%%%%
\newpage

%%%%%%%%%%%%%%%%%%%%%%%%%%%%%%%%%%%%%%%%%%%%%%
\section{Introduction}\label{Intro}    %  1  %
%%%%%%%%%%%%%%%%%%%%%%%%%%%%%%%%%%%%%%%%%%%%%%

Curved parallelizable manifolds are exceptional. For them, a connection can be defined on the frame bundle so that parallel transport on any closed loop brings every vector back to itself, a property also known as absolute parallelism or parallelism at a distance. 

In 1926, using tools of Riemannian geometry and group theory, E. Cartan and J. A. Schouten proved that $S^1$, $S^3$ and $S^7$ are parallelizable, that is, they admit flat $\mathfrak{u}(1)$, $\mathfrak{so}(3)$ and $\mathfrak{so}(7)$ connections, respectively \cite{Cartan-Schouten}. This result was later shown to be unique and true only for these spheres in a series of separate works by R. Bott, J. Milnor \cite{Bott-Milnor} and J. F. Adams \cite{Adams}, and building on results by H. Hopf \cite{Hopf} and A. Hurwitz \cite{Hurwitz} that the only normed division algebras have dimensions $1,2,4,8$ --corresponding to $\mathbb{R}$, $\mathbb{C}$, $\mathbb{H}$ and $\mathbb{O}$--, established that the only parallelizable spheres are $S^n$ with $n=0,1,3,7$ (see, e.g., \cite{Ranicki, Baez}).

The results of Cartan and Schouten contained some gaps in the proof of the case of $S^7$, which were addressed and resolved by J. A. Wolf, extending the result to the pseudo-Riemannian group manifolds and coset manifolds $S^{p,q}$ or $H^{p,q}$ with $p+q=7$ \cite{Wolf1,Wolf2}. Wolf's result established that any connected pseudo-Riemannian manifold which admits a flat connection is locally isomorphic to a globally symmetric pseudo-Riemannian manifold which is a product of a certain nilpotent Lie group manifold, flat space, group manifolds and seven-spheres.

The spheres $S^3$ and $S^7$ are manifolds of positive definite --Riemannian-- metric and it is therefore a natural question whether these results could be extended to manifolds of indefinite metric, like Lorentzian spacetimes or more generally, to manifolds of signature $(p,q)$.\footnote{Here the signature is defined by the metric structure on the tangent space $\eta_{ab}=diag (\overbrace{-,...,-}^{q},\overbrace{+,...,+}^{p})$, invariant under $SO(p,q)$. Reversing the spacetime signature ($\eta_{ab}\to -\eta_{ab}$) is a conventional change of which sign is associated to timelike or spacelike distances. The mirror conventions require consistent definitions of what is meant by positive and negative curvature. Here we will use the convention ''mostly plus'' which, for odd-dimensional spacetimes, gives $det(\eta)=-1$.}\label{Footnote_1} 

In a $D$-dimensional parallelizable manifold an orthonormal basis at any point $x$ can be parallel-transported to any other point $x'$ in a consistent manner, independently of the path taken to connect $x$ to $x'$. Hence, these manifolds admit $D$ linearly independent, globally defined vector fields. 

One can expect that some highly symmetric manifolds such as the pseudo-spheres $S^{p,q}=SO(p+1,q)/SO(p,q)$ with positive curvature, or pseudo-hyperbolic spaces $H^{p,q}=SO(p,q+1)/SO(p,q)$ with negative curvature, could be parallelizable as well. Here we present a constructive approach to find the flat $\mathfrak{so}(p,q)$-connections for $p+q=3,7$ and $p-q=1$. Our analysis leads to their explicit forms in the following cases:
\begin{itemize}
\item
In three dimensions, $AdS_3=H^{2,1}=SO(2,2)/SO(2,1)$ with signature (2,1) \cite{Alvarez:2014uda}. 
\item
In seven dimensions, $H^{4,3}=SO(4,4)/SO(4,3)$  with signature (4,3) \cite{Wolf1,Wolf2}.
\end{itemize}
In all of these cases torsion is necessarily present. The reversed signature manifolds $S^{1,2}=SO(2,2)/SO(1,2)$ and $S^{3,4}=SO(4,4)/SO(3,4)$ are also parallelizable. Following the comment in Footnote \ref{Footnote_1}, we will not discuss them any further throughout the text.

The occurrence of spheres with non-Euclidean signatures in different aspects of supergravity makes a detailed study of them an interesting problem for which our construction might be useful. It is known that the seven-sphere $S^7$ plays an important role in supergravity where the maximal supergravity in $4$-dimensions in anti-de Sitter (AdS$_4$) background with a local $SO(8)$ symmetry is obtained by the dimensional reduction of eleven-dimensional supergravity on $S^7$ \cite{deWit:1981sst,deWit:1982bul}. The seven-sphere $S^7$ has four coset representations $SO(8)/SO(7)$, $SO(7)/G_2$, $SO(6)/SU(3)$ and $SO(5)/Sp(2)$, where three latter ones are called squashed $S^7$ and each has a different holonomy group. It is known that one can also consider the dimensional reduction on squashed spheres \cite{Castellani:1983yg,Englert:1982vs}, in which case the resulting $4$-dimensional supergravity theory will preserve fewer supersymmetries, which can be understood as due to smaller holonomy groups \cite{Friedrich:1997} of the squashed spheres. For example the compactification on $S^7=Sp(4)/Sp(2)$ will lead to $\mathcal{N}=1$ or $\mathcal{N}=2$ supergravities \cite{Castellani:1983yg} and on $S^7=SO(7)/G_2$ the presence of internal fluxes leads to a supergravity theory in four dimensions with all supersymmetries broken \cite{Englert:1982vs}.

As discussed in \cite{Hull:1998fh,Hull:1999mt}, a similar role is also played by $H^{4,3}=SO(4,4)/SO(4,3)$ (or rather by $S^{3,4}=SO(4,4)/SO(3,4)$) as an internal manifold in the compactification of timelike T-dual M-theory (more precisely the supergravity theory at low energy limit of $M'$-theory) \cite{Hull:1998ym}, where one obtains an AdS$_4$ maximal supergravity with a local $SO(4,4)$ symmetry. An extension of this analysis in \cite{Henneaux:2017afd} showed that the squashed $H^{4,3}= SO(4,3)/G_{2,2}$ can be realized as an internal manifold in the dimensional reduction of $M'$-theory. In that case, the resulting $4$-dimensional supergravity is a non-supersymmetric theory.

The paper is organized as follows. In Section \ref{II} we show that covariantly constant torsion is a necessary condition for a manifold to have a flat connection. We then review the previous results by Cartan and Schouten \cite{Cartan-Schouten} and their generalization by Wolf \cite{Wolf1, Wolf2} to pseudo-Riemannian metric. In Section \ref{III} we introduce $\mathfrak{so}$-flatness which is relevant of our analysis for sphere or hyperbolic manifolds. We exhibit the idea explicitly and the role of torsion in the construction. Starting with three-dimensions we point out the need of  cross product and the relation with the quaternion and octonion division algebras (and their split forms). We also discuss the path along which it can be generalized to seven-dimensions. In Section \ref{sec:three_to_seven} we write down the flat connection both in the case of metric with Euclidean and non-Euclidean signature and identify the corresponding coset manifold in each case. We summarize the results in Section \ref{sec:discussion} and we include an appendix with a short review of division algebras and how they relate to parallelizability of spheres.

 %%%%%%%%%%%%%%%%%%%%%%%%%%%%%%%%%%%%%%%%%%%%%%%%%%%%%%%
\section{Previous results and notation}\label{II} % 2  %
%%%%%%%%%%%%%%%%%%%%%%%%%%%%%%%%%%%%%%%%%%%%%%%%%%%%%%%%

Consider a manifold $M^{p,q}$ equipped with a metric $g_{\mu \nu}(x)$, whose tangent space at each point $T_x$ is flat and equipped with a constant flat metric $\eta_{ab}$ invariant under $SO(p,q)$. The metric structure in $M^{p,q}$ is provided by the vielbein $e^a=e^a_\mu(x) dx^\mu$,
\begin{equation}
    g_{\mu \nu}(x)= \eta_{ab} e^a_\mu(x) e^b_\nu(x).
\end{equation}
The group $SO(p,q)$ acts locally on $M^{p,q}$ so that a vector in the tangent space $u^a(x)$ at a given point $x$ is parallel-transported to a nearby point $x+dx$ into $u^a(x+dx)= u^a(x)- dx^\mu \omega^a{}_{b\mu}(x)u^b(x)$, where $\delta^a{}_b + \omega^a{}_{b\mu}(x) dx^\mu$ is an element of $SO(p,q)$ infinitesimally close to the identity. This condition is also  expressed as
\begin{equation} \label{Du=0}
 Du^a \equiv dx^\mu[\partial_\mu u^a + \omega^a{}_{b\mu}(x) u^b]=0 \, ,
\end{equation}
where $Du^a$ is the covariant derivative of $u^a$ with connection 1-form $\omega^a{}_b = \omega^a{}_{b\mu} dx^\mu$ that takes values in the Lie algebra $\mathfrak{so}(p,q)$. 

The manifold $M^{p,q}$ is parallelizable if there exists a flat $SO(p,q)$ connection, namely, if the corresponding curvature two-form $R^a{}_b$, defined as\footnote{From now on wedge products ($\wedge$) of exterior forms will be implicitly understood.}
\begin{equation}\label{R}
R^a{}_b =d\omega^a{}_b +\omega^a{}_c\, \omega^c{}_b\, .
\end{equation}
vanishes identically on $M^{p,q}$. In what follows, we will refer to a connection for the $SO(p,q)$ group as an ``$\mathfrak{so}$-connection".

Another important feature of $M^{p,q}$ is its torsion two-form,
\begin{equation} \label{T}
    T^a= d e^a + \omega^a{}_b\, e^b\, ,
\end{equation}
which is in general independent of the curvature $R^a{}_b$. The covariant derivative of torsion, however, satisfies the identity
\begin{equation} \label{DT}
    DT^a \equiv R\indices{^a_b}\, e^b.
\end{equation}

%%%%%%%%%%%%%%%%%%%%%%%%%%%%%%%%%%%%%%%
\subsection{$\mathfrak{so}$-flatness}
%%%%%%%%%%%%%%%%%%%%%%%%%%%%%%%%%%%%%%%

From (\ref{DT}) it follows that torsion in a $\mathfrak{so}$-flat manifold need not vanish but must be covariantly constant, 
\begin{equation}\label{DT=0}
DT^a=0\, .
\end{equation}
In fact, as we will see, $\mathfrak{so}(p,q)$-flatness of a $(p+q)$-dimensional constant curvature manifold requires $T^a\neq 0$. Consider the torsion two-form expressed in a basis of local orthonormal frames as
\begin{equation}\label{T2}
T^a =\tau^a{}_{bc} \; e^b e^c \, ,
\end{equation}
where $\tau^a{}_{bc}$ is an $\mathfrak{so}$-tensor zero-form, antisymmetric in the lower indices. Then, a necessary condition for $\mathfrak{so}$-flatness (\ref{DT=0}) is
\begin{equation} \label{Gen-sol}
(D \tau^a{}_{bc}) e^b e^c + 2\tau^a{}_{bc}\tau^b{}_{df} e^c e^d e^f  =0\, ,
\end{equation}
which is satisfied if the coefficients multiplying $(e^a e^b)$ and $(e^a e^b e^c)$ vanish independently,
\begin{align} \label{Suff-cond-a}
D \tau^a{}_{bc} =0, \\ \label{Suff-cond-b}
\tau^a{}_{b[c}\tau^b{}_{df]}=0.
\end{align}

Condition (\ref{Suff-cond-a}) implies that $\tau^a{}_{bc}$ must be covariantly constant and if the connection is flat it can always be chosen as constant in an open region. Hence, without loss of generality one can take $d\tau^a{}_{bc}=0$, so that in the right basis, $\tau^a{}_{bc}$ is a constant tensor. As we shall see in the next section, conditions \eqref{Suff-cond-a} and \eqref{Suff-cond-b} are easily achieved in three dimensions; the situation in seven dimensions is less straightforward but non-trivial solutions still exist.

%%%%%%%%%%%%%%%%%%%%%%%%%%%%%%%%%%%%%%%
\subsection{Torsion free $\mathfrak{so}$-connection}
%%%%%%%%%%%%%%%%%%%%%%%%%%%%%%%%%%%%%%%

By definition, in any open chart of $M^{p,q}$ the covariant derivative of $\eta_{ab}$ vanishes identically, $D\eta^{ab} = 0$ \cite{Schutz,GS}. A necessary and sufficient condition for $D\eta^{ab} = 0$ is that the $\mathfrak{so}$-connections be antisymmetric,
\begin{equation} \label{omega-ab}
\omega_{ab}=-\omega_{ba}\; , \;\;\; \mbox{where} \;\;\;  \omega_{ab}:=\eta_{ac} \omega^c{}_b\, .
\end{equation}
A consequence of this is that if $\kappa_{ab}$ is any antisymmetric $SO(p,q)$-tensor one-form and $\omega^a{}_b$ is an $\mathfrak{so}$-connection, then $\omega^a{}_b + \kappa^a{}_b$ is also an $\mathfrak{so}$-connection. This freedom is usually associated to the possibility of choosing a torsion-free connection and, as shown below, it can also be used to construct flat $\mathfrak{so}$-connections.

The question we would like to address here is what are the conditions for a pseudo-Riemannian manifold $M^{p,q}$ to admit absolute parallelism. The proof of Cartan-Schouten as well as those of Bott-Milnor and Adams apply to Euclidean signature ($M^{p,q}=M^{n,0}\equiv M^{n}$), concluding that the only parallelizable spheres are $S^1$, $S^3$ and $S^7$. However, it is easy to show that AdS$_3=H^{2,1}$, the three-dimensional manifold of constant negative curvature and signature $(-,+,+)$, also admits a flat connection \cite{Alvarez:2014uda}. Here we examine how this result could extend to higher dimensions. Seven-dimensional constant curvature manifolds are obviously good candidates, but even though a priori there are several options for the possible signatures, it will be shown that only two different signatures are admissible up to the discrete symmetry $\eta_{ab} \rightarrow -\eta_{ab}$.

As is well known, it is possible to choose a particular $\mathfrak{so}$-connection $\bar{\omega}$ that satisfies the additional constraint of vanishing torsion,
\begin{equation} \label{omega-bar}
de^a + \bar{\omega}^a{}_b e^b \equiv 0 \; .
\end{equation}
This relation is algebraically solved for the torsion-free connection which is entirely determined by the metric structure,
\begin{equation} \label{omega-gamma}
\bar{\omega}^a{}_b = e^a{}_\nu (\Gamma\indices{^\nu_\mu} E\indices{^\mu_b} + d E\indices{^\nu_b})\,\, ,
\end{equation}
where $E\indices{^\mu_a}$ is the inverse vielbein and $\Gamma^\nu{}_\mu=\Gamma^\nu{}_{\mu\lambda} dx^\lambda$ is the Levi-Civita connection one-form, $\Gamma\indices{^\nu_{\mu\lambda}} = g^{\nu \gamma}(g_{\gamma \lambda},_\mu + g_{\gamma \mu},_\lambda-g_{\lambda \mu},_\gamma)$. Relation (\ref{omega-gamma}) can also be interpreted as relating the connections $\Gamma$ and $\bar{\omega}$ by a ``gauge transformation" represented by the Jacobian $e^a{}_\mu = (\partial z^a /\partial x^\mu)$ and its inverse $E\indices{^\mu_a}$. The reciprocal relation gives the Levi-Civita connection in terms of the torsion-free $\mathfrak{so}$-connection, 
\begin{equation} \label{gamma-omega}
\Gamma^\nu{}_\mu = E^\nu{}_a\,(\bar{\omega}^a{}_b e\indices{^b_\mu} + d e\indices{^a_\mu}) \, .
\end{equation}
This equivalence between the Levi-Civita connection and torsion-free $\mathfrak{so}$-connection is assumed to be globally valid by virtue of the invertibility of $e\indices{^a_\mu}$.\footnote{The invertibility of the vielbein might fail in the presence of topological defects in spacetime of the D'Auria-Regge type \cite{D'A-R}.} 

The Riemann curvature two-form obtained from the Levi-Civita connection is
\begin{equation} \label{calR}
\mathcal{R}\indices{^\alpha_\beta} =d \Gamma\indices{^\alpha_\beta} + \Gamma\indices{^\alpha_\gamma} \Gamma\indices{^\gamma_\beta}\, .
\end{equation}
This curvature is similar to --but quite different from-- the curvature (\ref{R}) defined by the $\mathfrak{so}$-connection $\omega^a{}_b$: $\Gamma^\alpha{}_\beta$ acts on tensors in a coordinate basis, while $\omega^a{}_b$ acts on tangent space tensors referred to an orthonormal basis. As stated in (\ref{omega-ab}) $\omega_{ab}=\eta_{ac}\,\omega^c{}_b$ is antisymmetric in its lower indices ($a \leftrightarrow b$), whereas $g_{\alpha \gamma}\Gamma^\gamma{}_\beta$ has no symmetry under exchange of $\alpha$ and $\beta$. More importantly, $\Gamma^\alpha{}_\beta$ is entirely defined by the metric of the manifold $g_{\mu \nu}$, while $\omega^a{}_b$ is independent from the metric structure. The two $\mathfrak{so}$-connections $\omega^a{}_b$ and $\bar{\omega}^a{}_b$ define different notions of parallelism and different $\mathfrak{so}$-curvatures, $R^a{}_b$ and $\bar{R}^a{}_b$, respectively. Their difference,
\begin{equation}
\omega^a{}_b -\bar{\omega}^a{}_b = \kappa^a{}_b.
 \end{equation} 
is the contorsion tensor, and the corresponding $\mathfrak{so}$-curvatures are related by
\begin{equation} \label{R-R bar}
R^a{}_b = \bar{R}^a{}_b + \bar{D}\kappa^a{}_b + \kappa^a{}_c\kappa^c{}_b\; ,
\end{equation}
where $ \bar{D}$ is the derivative for the $\bar{\omega}$ connection. From (\ref{gamma-omega}) the Riemann curvature $\mathcal{R}$ and the torsion-free $\mathfrak{so}$-curvature $\bar{R}$ are found to be linearly related,
\begin{equation} \label{calR-Rbar}
\mathcal{R}^\alpha{}_\beta = E^\alpha{}_a\bar{R}^a{}_b e^b{}_\beta \;, \end{equation}
and for example, $\mathcal{R}^\alpha{}_\beta =0 \Leftrightarrow \bar{R}^a{}_b=0$. 

Consider the Euclidean 3-sphere of radius $\rho$, defined as the surface $(x^1)^2+(x^2)^2+(x^3)^2+(x^4)^2 = \rho^2$ embedded in $\mathbb{R}^4$. Its Riemann curvature is
\begin{equation} \label{R-forS3}
\mathcal{R}^{\alpha \beta}{}_{\mu \nu} = (\delta^\alpha_\mu \delta^\beta_\nu - \delta^\alpha_\nu \delta^\beta_\mu)\rho^{-2}\;, \; \mbox{ or}\;\;\; \bar{R}^a{}_b=\rho^{-2} e^a e_b  \; .
\end{equation}
Hence, from (\ref{calR-Rbar}) one concludes that the torsion-free $\mathfrak{so}(3)$-connection $\bar{\omega}$ for $S^3$ is not flat. Parallelizability of $S^3$ means that there exist some other $\mathfrak{so}(3)$-connection $\omega$ whose curvature vanishes, but the corresponding torsion does not. 

Before going any further let us clarify an important point. Consider an $\mathfrak{so}(p,q)$ connection $\omega$ in a $(p+q)$-dimensional constant curvature manifold $M$ equipped with a local orthonormal basis $e^a$. It can be shown that $M$ admits a flat $\mathfrak{so}(p+1,q)$ or $\mathfrak{so}(p,q+1)$ connection: The one-form
\begin{equation} \label{W}
W^A{}_B = \left(\begin{array}{cc}
    \omega^a{}_b & e^a/\ell \\
     -\sigma e^b/\ell & 0
\end{array} \right) \, ,
\end{equation}
where $\ell$ is a constant, defines a connection for $\mathfrak{so}(p+1,q)$ (if $\sigma=1$) or $\mathfrak{so}(p,q+1)$ (if $\sigma=-1$). Hence, the curvature $F^A{}_B=dW^A{}_B + W^A{}_C W^C{}_B$ is
\begin{equation} \label{F}
F^A{}_B= \left( \begin{array}{cc}
R^a{}_b -\sigma e^a e_b/\ell^2& T^a/\ell \\
-\sigma T_b/\ell & 0
\end{array} \right) .
\end{equation}
Consequently, if $M$ is a torsion-free manifold of constant curvature $\sigma/\ell^2$ it admits a $\mathfrak{so}(p+1,q)$- or $\mathfrak{so}(p,q+1)$ flat connection $W$ whose curvature $F$ vanishes. It should be stressed that this is a feature of any constant curvature torsion-free manifold and independent from the statement that $S^3$ and $S^7$ admit flat $\mathfrak{so}(3)$ and $\mathfrak{so}(7)$ connections, respectively. For example, according to the above argument $S^4$ admits a flat $\mathfrak{so}(5)$-connection but it \emph{does not} admit a flat $\mathfrak{so}(4)$-connection and therefore is not said to be parallelizable.

%%%%%%%%%%%%%%%%%%%%%%%%%%%%%%%%%%%%%%%%%%%%%%%%%%%%%%
\section{Three dimensions}\label{III}  %  3  %
%%%%%%%%%%%%%%%%%%%%%%%%%%%%%%%%%%%%%%%%%%%%%%%%%%%%%%

In three dimensions, conditions \eqref{Suff-cond-a} and \eqref{Suff-cond-b} are identically satisfied by $\tau^a{}_{bc} \equiv \tau\epsilon^a{}_{bc}$ where $\epsilon_{abc}$ is the Levi-Civita tensor\footnote{Here $\epsilon_{abc}$ is a completely antisymmetric tensor in the tangent space with the convention $\epsilon_{012}=+1$. The tangent space indices $a,b,c,...$ are raised and lowered with $\eta^{ab}$ and $\eta_{ab}$.} and $\tau$ is a constant. In this case, $\tau^a{}_{bc}$ is an invariant tensor of $SO(p,q)$, ($p+q=3$) and (\ref{Suff-cond-b}) is just the Jacobi identity. Consequently, in three-dimensions the torsion two-form can be written as 
\begin{equation}\label{T= epsilon ee}
T^a= \tau \epsilon^a{}_{bc} e^b e^c \,.
 \end{equation}
The contorsion is $\kappa^a{}_b =-\tau \epsilon^a{}_{bc} e^c$, and
\be
\bar{D}\kappa\indices{^a_b}=0,\qquad  \mbox{and}\qquad  \kappa^a{}_c \kappa^c{}_b = \tau^2\, det(\eta)\, \delta^a_f\, \eta_{bd}\, e^d e^f \, ,
\ee
where $\bar{D}$ is the covariant derivative corresponding to the torsion-free connection $\bar{\omega}$, and we have used $\epsilon\indices{^a_{cd}}\,\epsilon\indices{^c_{bf}} = det(\eta)\, \left[\delta^a_f\, \eta_{bd} - \delta^a_b\, \eta_{df}\right]$. Hence, \eqref{R-R bar} reads
\begin{equation}\label{R=R + tee}
R^{ab} = \bar{R}^{ab} - \tau^2\, det(\eta)\, e^a e^b \, .
\end{equation}
Hence, the torsion-free curvature of a three-dimensional $\mathfrak{so}$-flat manifold is
\be
\bar{R}^a{}_b = \tau^2\, det(\eta)\, \eta_{bd}\, e^a e^d \, .
\ee
Conversely, this result implies that in a three-dimensional constant curvature manifold of radius $\rho$, the connection $\omega^a{}_b = \bar{\omega}^a{}_b - \rho^{-1}\epsilon^a{}_{bc} e^c$ is $\mathfrak{so}$-flat for the appropriate sign of $det(\eta)$. Hence, there are two distinct possibilities for a three-dimensional manifold admitting an $\mathfrak{so}$-flat connection:

\begin{itemize}
\item If the $\mathfrak{so}$-invariant metric $\eta_{ab}$ is Euclidean, i.e. $\eta_{ab}=diag(+1,+1,+1)$, the three-dimensional manifold has constant positive curvature, i.e. $S^3$, as expected.

\item If the $\mathfrak{so}$-invariant metric $\eta_{ab}$ is Lorentzian, $\eta_{ab}=diag(-1,+1,+1)$, then the manifold has constant negative curvature, i.e. AdS$_3$, \cite{Alvarez:2014uda}.
\end{itemize}

There is also the degenerate case $\tau = 0$ that corresponds to flat Euclidean and Lorentzian manifolds $\mathbb{R}^3$ and $\mathbb{R}^{2,1}$.

The key point in the previous construction is the possibility of writing the torsion two-form $T^a$ as the product (\ref{T2}), where $\tau^a{}_{bc}$ is an $\mathfrak{so}$-tensor zero-form whose function is essentially to map two vectors into a third one. In other words, $\tau$ defines an antisymmetric bilinear map from the tangent space $T{\mathcal{M}}$ onto itself, namely, a cross product:
\be \label{x-product}
\tau:\;\; T{\mathcal{M}} \times T{\mathcal{M}} \rightarrow T{\mathcal{M}}\, .
\ee

In three dimensions (\ref{T= epsilon ee}) identifies $\tau^a_{~bc}$ with $\epsilon^a_{~bc}$, the invariant $\mathfrak{so}$-tensor that defines the cross product of three-dimensional vectors (up to a constant). In addition, $\epsilon^a{}_{bc}$ also defines the structure constants of the Lie algebra $\mathfrak{so}(3)$ --or $\mathfrak{so}(2,1)$--, which is the feature behind the fulfillment of conditions \eqref{Suff-cond-a} and \eqref{Suff-cond-b}. 

It can be seen that the existence of the cross product in 3-dimensional Euclidean space $V^3$ is related to the isomorphism with the imaginary quaternions (see Appendix),
\begin{equation}
    V^3 \cong \textrm{Im}\, \mathbb{H}
\end{equation}

The fact that the construction also works for the three-dimensional Lorentzian space $ V^{2,1}$ is due to the existence of another cross product $\times_s$ which can be traced back to another isomorphism, the one between $ V^{2,1}$ and the imaginary \textit{split} quaternions, 
\begin{equation}
V^{2,1} \cong  \textrm{Im}\, \mathbb{H}_s.
\end{equation}

The extension of these ideas to seven dimensions is rather straightforward, as we discuss in the next section.
 %%%%%%%%%%%%%%%%%%%%%%%%%%%%%%%%%%%%%%%%%%%%%%%%%%%%%%%%%%%%%%%%%
\section{Seven dimensions}\label{sec:three_to_seven} % 4 %
%%%%%%%%%%%%%%%%%%%%%%%%%%%%%%%%%%%%%%%%%%%%%%%%%%%%%%%%%%%%%%%%%%
In seven-dimensional vector spaces there also exist two antisymmetric bilinear maps, the cross products $\times$ and $\times_s$. The first results from the isomorphism between $V^{7}$ and the imaginary octonions, 
\begin{equation}
    V^{7} \cong \textrm{Im}\,\mathbb{O} ,
\end{equation}
and the second follows from the isomorphism between $V^{4,3}$ and the imaginary split octonions, 
\begin{equation}
    V^{4,3}\cong \textrm{Im}\, \mathbb{O}_s.
\end{equation}

In the first case, the cross product $\times$ is provided by the tensor $f_{abc}$ that defines the multiplication rule of the octonions, see \eqref{o-lambdas} in Appendix. In the second case the cross product $\times_s$ is brought in by the tensor $\tilde{f}_{abc}$ that defines the multiplication rule of the split-octonions, see \eqref{split-product} in Appendix. We will next review these two cases separately.

%%%%%%%%%%%%%%%%%%%%%%
\subsection{Euclidean signature} %  4.1 %
%%%%%%%%%%%%%%%%%%%%%%

Unlike the situation in three dimensions, where the structure constants $\epsilon^a{}_{bc}$ define an invariant $SO(3)$ tensor, the coefficients $f^a{}_{bc}$ transform as the components of an $SO(7)$ tensor, but is not as an \textit{invariant} $SO(7)$ tensor. The coefficients $f^a{}_{bc}$ define an invariant tensor under the exceptional Lie group $G_2 \subset SO(7)$, the automorphism group of the octonions \cite{Humphreys2012}. 

In fact, $G_2$ can be defined as the automorphism group of $f_{abc}$ and therefore it is also the group of real linear transformations that preserve the cross product. Additionally, $G_2$ is also the isotropy group of $Spin(7)$,\footnote{$Spin(7)$ is the double cover of $SO(7)$ and can be built with the use of spinor representations of $\mathfrak{so}(7)$.} which leaves the identity of octonions fixed and therefore preserves the space orthogonal to the identity, the seven-dimensional space of imaginary octonions $\textrm{Im}\,\mathbb{O}$. Finally, since $G_2$ is a subgroup of $SO(7)$, there is an inclusion $G_2 \hookrightarrow{} SO(\textrm{Im}\,\mathbb{O})$ and therefore $G_2$ admits a 7-dimensional representation $\textrm{Im}\,\mathbb{O}$.

The exceptional group $G_2$ is the subgroup of $SO(7)$ which preserves the non-degenerate $3$-form \cite{Bryant1987, Agricola2008}
\be
\varphi= dx^{123}+dx^{145}+dx^{176}+dx^{246}+dx^{257}+dx^{347}+dx^{365} = \frac{1}{3!}f_{abc}dx^{abc},
\ee
where $dx^{abc}= dx^a \wedge dx^b \wedge dx^c$ \cite{Baez,Ranjbar:2018hvy}.

Since the coefficients $f_{abc}$ define an invariant tensor under $G_2 \subset SO(7)$ it means that for any group element $\Lambda^a{}_b \in G_2$ , the coefficients\footnote{We assume Euclidean signature so that $f^a{}_{bc}$ takes the same values as $f_{abc}$, but we still keep track of upper and lower indices to facilitate the transition to other signatures.} $f^a{}_{bc}$ transform as the components of an invariant tensor, i.e. $\Lambda:f \rightarrow f$, or
\be \label{g2-invariant}
\Lambda f:= \Lambda\indices{^a_d} f\indices{^d _{eh}} (\Lambda^{-1})\indices{^e_b}(\Lambda^{-1})\indices{^h_c}=f\indices{^a_{bc}}.
\ee
This guarantees that $G_2$ is an automorphism of $ \textrm{Im}\, \mathbb{O}$ for the cross product,
\begin{equation}\label{g-cross-product}
\Lambda\in G_2 \Longrightarrow \Lambda(v^a \times v^b) = \Lambda(v^a) \times \Lambda(v^b).
\end{equation}
Let $\Lambda\in G_2$ be an element near the identity, 
\begin{equation}
\Lambda^a{}_b= \delta^a_b + \theta^s (\mathbb{J}_s)^a{}_b, 
\end{equation}
where the generators $\mathbb{J}_s\, (s=1,2,\cdots,14)$ belong to $\mathfrak{g}_2$, the Lie algebra of $G_2$, satisfying
\begin{equation}\label{g2}
[\mathbb{J}_r, \mathbb{J}_s] = C^t_{rs}\,  \mathbb{J}_t\; ,
\end{equation}
where $C^t_{rs}$ are the structure constants of $\mathfrak{g}_2$. 

From (\ref{g2-invariant}), the invariance of the cross product under the action of $G_2$ implies
\begin{equation}\label{Jf-Jf-Jf=0}
(\mathbb{J}_s)^a{}_d f^d{}_{bc} - (\mathbb{J}_s)^d{}_b f^a{}_{dc} - (\mathbb{J}_s)^d{}_c f^a{}_{bd} = 0.
\end{equation}

Here $\{\mathbb{J}_s\}$ are the 14 generators of $G_2$ in a $7\times7$ antisymmetric representation of the Lie algebra $\mathfrak{g}_2$ (for a concrete example of this  representation, see \cite{Gunaydin-Gursey}). The group $SO(7)$ contains seven additional generators $\mathbb{X}_1, \cdots, \mathbb{X}_7$ which can be chosen as 
\begin{equation}\label{J=f}
(\mathbb{X}_c)^a{}_b = f^a{}_{bc}.
\end{equation}
The commutator of $\mathbb{X}$ with the $G_2$ generators is
\begin{eqnarray}
\left[ \mathbb{J}_s, \mathbb{X}_c \right]^a{}_b &=& (\mathbb{J}_s)^a{}_d (\mathbb{X}_c)^d{}_b - (\mathbb{X}_c)^a{}_d (\mathbb{J}_s)^d{}_b \\
&=&  (\mathbb{J}_s)^a{}_d f^d{}_{bc} - (\mathbb{J}_s)^d{}_b f^a{}_{dc}= (\mathbb{J}_s)^d{}_c f\indices{^a_{bd}}  \\
&=& (\mathbb{J}_s)^d{}_c  (\mathbb{X}_d)^a{}_b \, ,
\end{eqnarray}
where (\ref{Jf-Jf-Jf=0}) was used to write the second line. This can be recognized as stating that the generators $\{\mathbb{X}_a\}$ transform as $G_2$ vectors,
\begin{equation}
\left[ \mathbb{J}_s, \mathbb{X}_c \right] = (\mathbb{J}_s)^d{}_c \mathbb{X}_d \, .
\end{equation}
 The generators $\mathbb{X}$ as  well as their commutators are antisymmetric $7 \times 7$ matrices and therefore they leave the seven-dimensional Euclidean metric $\delta_{ab}$ invariant. Thus, the  $\mathbb{X}$'s are $SO(7)$ generators as well. 

Now, in a seven-dimensional manifold with Euclidean signature one can consider three different connections, $\hat{\omega}$, $\omega$ and $\bar{\omega}$, corresponding to three distinct notions of parallelism: parallel transport under $G_2$, $SO(7)$ and the torsion-free $SO(7)$ connections, respectively. Each of these connections defines a corresponding covariant derivative, which we denote by $\hat{D}$, $D$ and $\bar{D}$. Since $G_2\subset SO(7)$, $\hat{\omega}$ can also be viewed as a piece of the connection of $SO(7)$, $\omega$, or of the torsion-free $SO(7)$-connection, $\bar{\omega}$. These three connections have different independent properties as compared in Table \ref{Tab:Oct},

\begin{table}
\begin{center}
\begin{tabular}{|c|c|l|}
\hline 
Group    & Connection & $\;\;$  Property \\
\hline 
%\hline 
$G_2$    &  $\hat{\omega}$ & $\hat{D} f^a{}_{bc} =0$  \\ 
$SO(7)$  &  $\omega$	   &   $T^a=De^a \neq 0$ \\ 
$SO(7)$	 & $\bar{\omega}$ & $\bar{D} e^a  =0$ \\
\hline
\end{tabular}
\caption{The three connections on the tangent space for a generic seven-dimensional manifold with Euclidean metric.} \label{Tab:Oct}
\end{center}
\end{table}

The $G_2$ connection $\hat{\omega}(x)$ can be expressed as
\begin{equation}
\hat{\omega}^a{}_b = \hat{\omega}^s(x) \left(\mathbb{J}_s\right)^a{}_b \; \, ,  s=1, \cdots, 14.
\end{equation}

Consider now the $SO(7)$ connection defined as
\begin{align} 
\omega^a{}_b &= \hat{\omega}^s (\mathbb{J}_s)^a{}_b + \alpha e^c (\mathbb{X}_c)^a{}_b \nonumber\\
&= \hat{\omega}^a{}_b + \alpha f^a{}_{bc}e^c\, , \label{eq:omega-omegahat}
\end{align}
where $\alpha$ is a constant and $e^a$ is a vielbein one-form.

A relation analogous to (\ref{eq:omega-omegahat}) can be assumed between the $SO(7)$ connection $\omega$ and the torsion-free $SO(7)$ connection $\bar{\omega}$,
\begin{align}
\omega^a{}_b &= \bar{\omega}^s (\mathbb{J}_s)^a{}_b + \mu e^c (\mathbb{X}_c)^a{}_b \nonumber\\
&= \bar{\omega}^a{}_b + \mu f^a{}_{bc}e^c,\label{eq:rel_omega_omegabar}
\end{align}
where $\mu $  is another constant and $\bar{\omega}$ is defined by the torsion-free condition, $de^a + \bar{\omega}^a{}_b e^b =0$. Now the torsion is given by
\begin{equation} \label{T=-mufee}
T^a=De^a =  -\mu f^a{}_{bc}e^b e^c
\end{equation}
which, combined with (\ref{eq:omega-omegahat}) implies 
\begin{equation}
\hat{D} e^a = (\alpha - \mu) f^a{}_{bc} e^b e^c \, ,
\end{equation}
or equivalently
\be\label{eq:rel_omegabar_omegahat}
\bar{\omega}\indices{^a_b}=\hat{\omega}\indices{^a_b} + (\mu - \alpha) f\indices{^a_{bc}} e^c.
\ee

Furthermore, the covariantly constant torsion, $DT^a=0$, implies
\be
DT^a= -\mu (3\alpha - 2\mu) f\indices{^a_{d[g}}f\indices{^d_{bc]}} e^g e^b e^c =0,
\ee
Since $f^a{}_{d[g}f^d_{bc]}$ does not vanish (octonions are non-associative), this in turn leads to an important relation between the parameters $\alpha$ and $\mu$,
\be\label{eq:alpha-mu-rel}
\alpha = \f{2}{3} \mu.
\ee

Using the equation \eqref{eq:rel_omega_omegabar} and the second Cartan equation we can write down the $SO(7)$ curvature as
\be\label{eq:R_Rbar-pre1}
R\indices{^a_b} = \bar{R}\indices{^a_b} + \mu \bar{D} f\indices{^a_{bc}} e^c + \mu^2 f\indices{^a_{dc}} f\indices{^d_{bg}} e^c e^g.
\ee
One can simplify this by writing
\be
\bar{D}f\indices{^a_{bc}} e^c = \hat{D}f\indices{^a_{bc}} e^c + (\alpha-\mu) \left( f\indices{^a_{dg}} f\indices{^d_{bc}} + f\indices{^a_{dc}} f\indices{^d_{gb}} + f\indices{^a_{db}} f\indices{^d_{cg}}\right) e^g e^c,
\ee
where we have used \eqref{eq:rel_omegabar_omegahat}. Equation \eqref{eq:R_Rbar-pre1} can now be written as
\begin{align}
R\indices{^a_b} = \bar{R}\indices{^a_b} + \mu\, \hat{D} f\indices{^a_{bc}} e^c  &+ \mu^2 f\indices{^a_{dc}} f\indices{^d_{bg}} e^c e^g \nonumber\\
&+ \mu (\alpha-\mu) \left( f\indices{^a_{dg}} f\indices{^d_{bc}} + f\indices{^a_{dc}} f\indices{^d_{gb}} + f\indices{^a_{db}} f\indices{^d_{cg}}\right) e^g e^c.
\end{align}

Given the property $\hat{D} f\indices{^a_{bc}}=0$ (see Table \ref{Tab:Oct}), and the condition \eqref{eq:alpha-mu-rel}, we have
\be
R\indices{^a_b} = \bar{R}\indices{^a_b} +\f{1}{3} \mu^2 \left(f\indices{^a_{dc}} f\indices{^d_{bg}} + f\indices{^a_{db}} f\indices{^d_{cg}}\right) e^c e^g.
\ee

Direct computation using the explicit form of $f\indices{^a_{bc}}$ given in \eqref{f} implies that 
\begin{equation} \label{Rbar-R'}
R^a{}_b = \bar{R}^a{}_b + \mu^2 e^a e_b .
\end{equation}
Therefore an $SO(7)$-flat seven manifold must have constant \textit{positive} Riemannian curvature of radius $\mu^{-1}$,
\begin{equation}\label{Riem-curv-final-sym}
\bar{R}^a{}_b = \mu^2 e^a e_b = \frac{9}{4}\alpha^2 e^a e_b ,
\end{equation}
which states the known fact that a seven sphere admits a flat $SO(7)$ connection \eqref{eq:rel_omega_omegabar}.

Similarly to the Riemann curvature, one finds that the $G_2$ curvature of an $SO(7)$-flat seven-dimensional manifold is 
\begin{equation}
\hat{R}^a{}_b =-\frac{2}{9}\mu^2 \left(2 f\indices{^a_{dc}} f\indices{^d_{bg}} +  f\indices{^a_{db}} f\indices{^d_{cg}}\right) e^c e^g \, .
\end{equation}

Once again by using \eqref{f}, one can directly check that the $G_2$ curvature of the $SO(7)$-flat sphere is nonzero,\footnote{This means that under parallel transport on a closed loop, a tangent vector comes back rotated by an element of $G_2$: $e^a\longrightarrow \Lambda^a{}_b\, e^b$, with $\Lambda^a{}_b \in G_2$.}
\begin{equation}
\hat{R}^a{}_b =\frac{2}{9}\mu^2(4e^a e_b + *f\indices{^a_{bcd}} e^c e^d) \,.
\end{equation}
where $*f\indices{^a_{bcd}} \equiv \frac{1}{6}\epsilon^a{}_{bcdefg}f^{efg}$.

From the expressions for the torsion two-form (\ref{T=-mufee}) and the curvature two-form \eqref{Riem-curv-final-sym}, it is easy to see that their tensor components in the appropriate basis, $T\indices{^a_{bc}}=-\mu f^a{}_{bc}$ and $\bar{R}^a{}_{bcd}$ are invariant under $G_2$.

%%%%%%%%%%%%%%%%%%%%%%%%
\subsection{Non-Euclidean signature}\label{sec:Pseudo_Riem_Sign} % 4.2 %
%%%%%%%%%%%%%%%%%%%%%%%%

We now consider a manifold with metric $\eta_{ab}$ invariant under $SO(p,q)$ in the tangent space. The constructive approach of the previous subsection can be repeated provided a cross product can be consistently defined by some $SO(p,q)$ tensor $\tilde{f}^a{}_{bc}$. Then, the existence of a covariantly constant torsion hinges on the existence of a certain tensor invariant under a subgroup of $SO(p,q)$. It turns out that this is true for $SO(4,3)$ --and for its mirror, $SO(3,4)$-- in which case there exists a cross product in seven dimensions, $\times_s$,
\begin{equation}
(v \times_s  w)^a =\tilde{f}^a{}_{bc} v_b w^c,\qquad \, v, w, v \times_s w \in \textrm{Im}\,\mathbb{O}_s,
\end{equation}
which is now an invariant operation under $SO(4,3)$, see Appendix for information on $\textrm{Im}\,\mathbb{O}_s$ and the form of $\tilde{f}\indices{^a_{bc}}$. A cross product $\times_s$ can be constructed in analogy to that for $V^{2,1}$ starting from the multiplication table for the imaginary split octonions (we use the subscript $s$ to emphasize this point). The $SO(4,3)$-tensor $\tilde{f}$ is invariant under an exceptional group known as ``split-$G_2$", a subgroup of $SO(4,3)$ also denoted as $G_{2,2}$, the group of automorphism of split octonions $\mathbb{O}_s$ \cite{Agricola2008,Baez2}. 

Echoing the discussion of the previous section, one can define parallelism with respect to the $\mathfrak{so}(4,3)$ connection $\omega$, the torsion-free $\mathfrak{so}(4,3)$ connection $\bar{\omega}$ and the $\mathfrak{g}_{2,2}$ connection $\tilde{\omega}$. Each of these connections has a corresponding covariant derivative --denoted by $D$ and $\bar{D}$ and $\tilde{D}$,  respectively-- with different geometric features, as summarized in the Table \ref{Tab:split_connection}.

\begin{table}[h!]
\begin{center}
\begin{tabular}{|c|c|l|}
\hline
Group  &  Connection  &  $\;\;$  Property \\
\hline 
%\hline 
$G_{2,2}$  &  $\tilde{\omega}$  &  $\tilde{D} \tilde{f}^a{}_{bc} =0$  \\ 
$SO(4,3)$ 	&  $\omega$	  &   $T^a=De^a \neq 0$ \\ 
$SO(4,3)$  &  $\bar{\omega}$  &  $\bar{D} e^a  =0$ \\

\hline
\end{tabular}
\end{center}
\caption{The three connections on the tangent space for a generic seven-dimensional manifold with non-Euclidean metric.} \label{Tab:split_connection}
\end{table}

Since $G_{2,2}\subset SO(4,3)$, the connection $\tilde{\omega}$ can also be viewed\footnote{It is important to stress that in fact, instead of $G_{2,2}$, one has to take into account the group $G^*_{2,2}$ which is a centerless group to be obtained from $G_2$ and is the stabilizer of $SO^+(4,3)$. See \cite{Ranjbar:2018hvy,Henneaux:2017afd} for details.} as a piece of the connection of $SO(4,3)$. Hence, the generic $SO(4,3)$ connection $\omega$ can be written in a form similar to the previous section in terms of $\bar{\omega}$ and $\tilde{\omega}$ as
\be\label{eq:omega-omegatilde}
\omega\indices{^a_b}=\tilde{\omega}\indices{^a_b}+\tilde{\alpha} \tilde{f}\indices{^a_{bc}}e^c,
\ee
where the $\mathfrak{g}_{2,2}$ connection $\tilde{\omega}(x)$ can be expressed as
\begin{equation}
\tilde{\omega}^a{}_b = \tilde{\omega}^s(x) (\tilde{\mathbb{J}}_s)^a{}_b, \qquad  s=1, \cdots, 14
\end{equation}
and $\{\tilde{\mathbb{J}}_s\}$ are the 14 generators of $G_{2,2}$ in a $7\times7$ antisymmetric representation of the Lie algebra $\mathfrak{g}_{2,2}$. In analogy with (\ref{eq:rel_omega_omegabar}) we now write
\be\label{eq:flat_con_spin34_omega}
\omega\indices{^a_b}= \bar{\omega}\indices{^a_b} + \tilde{\mu} \tilde{f}^a{}_{bc} e^c,
\ee
and therefore torsion is written as
\be
T^a= -\tilde{\mu} \tilde{f}\indices{^a_{bc}} e^b e^c\,,
\ee

Since $\bar{\omega}\indices{^a_b}=\tilde{\omega}\indices{^a_b} + (\tilde{\alpha} - \tilde{\mu}) \tilde{f}\indices{^a_{bc}} e^c$,
the condition of covariantly constant torsion, $DT^a = 0$ leads again to the relation
\be
\tilde{\alpha} = \f{2}{3} \tilde{\mu}.
\ee
The corresponding curvatures $R$ and $\bar{R}$ are related by
\be\label{eq:curv_flat_spin34_con}
R\indices{^a_b} = \bar{R}\indices{^a_b} + \tilde{\mu}\,\bar{D} \tilde{f}\indices{^a_{bc}}\, e^c + \tilde{\mu}^2 \tilde{f}\indices{^a_{dc}} \tilde{f}\indices{^d_{bg}}\, e^c e^g \, ,
\ee
and 
\begin{align}
R\indices{^a_b} &= \bar{R}^a{}_b + \tilde{\mu}\, \tilde{D}\tilde{f}\indices{^a_{bc}}\, e^c + \tilde{\mu}^2 \tilde{f}\indices{^a_{dc}} \tilde{f}\indices{^d_{bg}}\, e^c e^g + \tilde{\mu} (\tilde{\alpha}-\tilde{\mu}) \left(\tilde{f}\indices{^a_{dg}} \tilde{f}\indices{^d_{bc}} + \tilde{f}\indices{^a_{dc}} \tilde{f}\indices{^d_{gb}}+ \tilde{f}\indices{^a_{db}} \tilde{f}\indices{^d_{cg}} \right) e^g e^c \nonumber\\
&= \bar{R}^a{}_b +\frac{\tilde{\mu}^2}{3} \left(\tilde{f}\indices{^a_{dc}} \tilde{f}\indices{^d_{bg}} - \tilde{f}\indices{^a_{bd}} \tilde{f}\indices{^d_{cg}} \right) e^c e^g 
\end{align}

Direct computation using the explicit form of $\tilde{f}^a{}_{bc}$ from the multiplication table \ref{Tab:Multipl-Split-Oct} yields
\be \label{R=barR-ee}
R^{ab} = \bar{R}^{ab}+\tilde{\mu}^2 e^a e^b.
\ee
Therefore an $SO(4,3)$-flat seven manifold must have constant \textit{negative} Riemannian curvature of radius $\tilde{\mu}^{-1}$,
\begin{equation}
\bar{R}^{ab} = -\tilde{\mu}^2 e^a e^b = -\frac{9}{4}\tilde{\alpha}^2 e^a e^b.
\end{equation}
This result states that a seven pseudo-hyperbolic space $H^{4,3}$ admits a flat $SO(4,3)$ connection given by \eqref{eq:flat_con_spin34_omega}.

%%%%%%%%%%%%%%%%%%%%%%%%%%%%%%%%%%%%%%%%%%%%%%%%%%%%%%
\section{Summary and discussion}\label{sec:discussion}
%%%%%%%%%%%%%%%%%%%%%%%%%%%%%%%%%%%%%%%%%%%%%%%%%%%%%%

We have studied the condition for having a flat $\mathfrak{so}(p,q)$-connection in the presence of torsion in three and seven dimensions. The following table lists curved (in the Riemannian, torsion-free sense) maximally symmetric parallelizable manifolds of dimensions three, seven and beyond with their corresponding flat connections.

\begin{table}[h!]
\begin{center}
\begin{tabular}{|c|l|l|l|l|}
\hline
Manifold  & $\mathfrak{so}(p,q)$ & Tangent space metric $\eta_{ab}$ & Curvature $\bar{R}^a{}_b$ &  Connection $\omega^a{}_b$\\
\hline 
$S^3$  & $\mathfrak{so}(3)$ & diag$(1,1,1)$ & \;\; $\rho^{-2}e^a e_b$  &  $\bar{\omega}^a{}_b-\rho^{-1}\epsilon^a{}_{bc} e^c$  \\ 
$H^{2,1}=$ AdS$_3$ & $\mathfrak{so}(2,1)$ & diag$(-1,1,1)$ &  $-\rho^{-2}e^a e_b$  &  $\bar{\omega}^a{}_b-\rho^{-1}\epsilon^a{}_{bc} e^c$  \\ 
$S^7$ & $\mathfrak{so}(7)$ & diag$(1,1,1,1,1,1,1)$ &  \;\; $\rho^{-2}e^a e_b$  & $\bar{\omega}^a{}_b-\rho^{-1}f^a{}_{bc} e^c$  \\
$H^{4,3}$ & $\mathfrak{so}(4,3)$ & diag$(-1,-1,-1,1,1,1,1)$ &  $-\rho^{-2}e^a e_b$  & $\bar{\omega}^a{}_b-\rho^{-1}\tilde{f}^a{}_{bc} e^c$ \\  
$\mathbb{R}^{p,q}$ & $\mathfrak{so}(p,q)$ & diag$(-1,\cdots,-1,1,\cdots,1)$ & \;\;\;\;0  & \;\;\;\;\;\;\;0 \\ \hline
\end{tabular}
\end{center}
\caption{Fundamental parallelizable manifolds with non-Euclidean tangent space metric of dimensions $3,7$ and $p+q$, and their corresponding flat $\mathfrak{so}(p,q)$ connections. The radius of curvature $\rho$ is $\mu^{-1}$ and $\tilde{\mu}^{-1}$ for $S^7$ and $H^{4,3}$ respectively. %\bc{For all the manifolds in this table, manifolds obtained from reversing the signature of tangent space metric ($\eta_{ab}\rightarrow -\eta_{ab}$) are also parallelizable. These mirrored manifolds will have a flat $so(q,p)$-connection with the opposite sign curvature.}
} \label{Tab:summary}
\end{table}

\subsection*{Extensions}
Note that we have not ruled out the existence of parallelizable pseudo-Riemannian manifolds of other dimensions. Unlike the case of positive definite tangent space metric, where the theorems of Hurwitz \cite{Hurwitz} or Adams \cite{Adams} guarantee the existence of only four distinct normed division algebras, there is no such a clear argument for the pseudo-Riemannian case. Hence, based on the construction presented here one cannot claim that pseudo-spheres $S^{p,q}$ or pseudo-hyperbolic spaces $H^{p,q}$ with $p+q > 7$ are not parallelizable.

Other parallelizable manifolds can be constructed as direct products of these maximally symmetric ones and also by quotienting them by sets of commuting Killing vectors. In the latter case, the new manifolds may contain singularities, such as the case of the 2+1 black hole and the related conical singularities \cite{Alvarez:2014uda}, in which case parallelizability is restricted to open sets that do not contain singularities.

Under conformal (Weyl) transformations the metric and the vielbein are locally rescaled,
\begin{equation} \label{Weyl}
    g_{\mu \nu}(x) \rightarrow g'_{\mu \nu}(x) = \Omega^2(x) g_{\mu \nu}(x)\;,\;\;\;\;
    e^a(x) \rightarrow e'^a(x) = \Omega(x)e^a(x).
\end{equation}
This transformation does not affect the isotropy group of the tangent space and therefore \eqref{Weyl} must commute with the $\mathfrak{so}(p,q)$ symmetry in the tangent space. Hence, \eqref{Weyl} leaves the connection unchanged,
\begin{equation}
 \omega^a{}_b(x) \rightarrow \omega'^a{}_b(x) = \omega^a{}_b(x).
\end{equation} 
However, the torsion-free $\mathfrak{so}(p,q)$ connection whose dependence on the vielbein is given in \eqref{omega-gamma}, changes as
\begin{equation}
  \bar{\omega}_{ab}(x) \rightarrow  \bar{\omega}'_{ab}(x) = \bar{\omega}_{ab}(x) + (e_a E^\rho{}_b - e_b E^\rho{}_a)\partial_\rho \log \Omega\;,
    \end{equation}
or, $\bar{\omega}'_{ab} = \bar{\omega}_{ab} + (e_a \partial_b - e_b \partial_a)\log \Omega$. Consequently, the $\mathfrak{so}(p,q)$ curvature $R^a{}_b$ is unchanged under \eqref{Weyl} while the torsion-free part $\bar{R}^a{}_b$ changes inhomogeneously in a non-symmetric form \cite{Yau}. This means that a manifold conformally-related to an $\mathfrak{so}(p,q)$ flat one is also parallelizable. This applies, in particular, to squashed spheres obtained by conformal deformations of a parallelizable one. 
%\gc{Since the Riemann curvature of the squashed $S^7=SO(7)/G_2$ and $S^{3,4}=SO(3,4)/G_{2,2}$ has non-symmetric form,\footnote{\bc{Such a coset is known to be a non-symmetric coset \cite{Englert:1982vs,Biran:1982eg,DAuria:1982chj,Gunaydin:1995as} as it can be seen from the commutator of those elements of $\mathfrak{so}(7)$ which are not in $\mathfrak{g}_2$ \cite{VanNieuwenhuizen:1985be,Castellani:1983tb,Castellani:1999fz}. In a symmetric coset, for $a,b \in \mathfrak{so}(7)/\mathfrak{g}_2$ we have $[a,b]\in \mathfrak{g}_2$ while in a non-symmetric coset such a commutator has a piece which belongs to the coset $\mathfrak{so}(7)/\mathfrak{g}_2$. The explicit formula for the curvature of a coset can be found in \cite{Castellani:1999fz}. }} it is suggestive that the squashed $S^7$ or $S^{3,4}$ are parallelizable as it has been shown in \cite{Friedrich:1997, Baez, Kath:1998}.}   \rc{[Do you know a ref. where the metric of $S^7=SO(7)/G_2$ can be found?]}

From the viewpoint of compactification in supergravity, the internal fluxes in the case of $S^7=SO(7)/G_2$ are components of torsion in the language of differential geometry. As a result, this manifold sometimes called round seven-sphere with torsion. As shown here, torsion is proportional to $f^a{}_{bc}$, which is invariant under $G_2 \subset SO(7)$ and therefore the symmetry group of the coset manifold breaks down from $SO(8)$ to $SO(7)$. This point can be seen with the observation that the imaginary octonion vector space is isomorphic to the tangent space of $S^7$. More precisely, the octonions form an eight-dimensional vector space that can be decomposed into the seven-dimensional space of imaginary octonion and the identity element which are perpendicular to each other. The group $G_2$ is the isotropy group of $Spin(7)$ which fixes the identity and since it is the automorphism group of octonions, it  thus preserves the space orthogonal to the identity. The imaginary octonion basis span a seven-dimensional vector space isomorphic to the tangent space of $S^7$. Thus, the seven-sphere is written as the coset $Spin(7)/G_2$.

Extending the proof to the pseudo-Riemannian case, we saw that a similar construction is possible for $H^{4,3}=SO(4,3)/G_{2,2}$ with torsion proportional to $\tilde{f}^a{}_{bc}$, which is invariant under $G_{2,2} \subset SO(4,3)$. In this case, torsion breaks the $SO(4,4)$ symmetry of round seven-pseudo-sphere (or seven-pseudo-hyperbolic space) down to $SO(4,3)$ \cite{Henneaux:2017afd}. This can also be seen geometrically using the isomorphism between the imaginary split octonion vector space and the tangent space on $H^{4,3}$.

Two comments are in order:
\begin{itemize}
\item 
The previous results can be expressed for metrics with both signatures. For example in expressions (\ref{Rbar-R'}) for $S^7$, and (\ref{R=barR-ee}) for $H^{4,3}$, the relative sign of the second term changes with the sign of $\det(\eta)$ and therefore one can write
\be \label{R=barR-pmee}
R^{ab} = \bar{R}^{ab}- \mu^2 det(\eta)\, e^a e^b \, ,
\ee
which is valid for all cases \cite{Ranjbar:2018hvy}.
\item Eq. \eqref{R=barR-pmee} means that given a three- or seven-dimensional manifold of constant Riemannian curvature, one can construct the corresponding flat $\mathfrak{so}$-connection, where $\mathfrak{so}$ stands for $\mathfrak{so}(3)$, $\mathfrak{so}(2,1)$, $\mathfrak{so}(7)$, $\mathfrak{so}(4,3)$.
\end{itemize}

We can summarize the result in seven dimensions as follows:

\begin{itemize}
\item If $T\mathcal{M}\cong \textrm{Im}\,\mathbb{O}$, then the induced metric $\eta_{ab}$ on the tangent space is Euclidean. The manifold is a seven dimensional constant positive curvature symmetric manifold, that is $S^7=\f{SO(8)}{SO(7)}$ or constant positive curvature non-symmetric coset space $S^7=\f{SO(7)}{G_2}$ in the presence of torsion.
\item If $T\mathcal{M}\cong \textrm{Im}\,\mathbb{O}_s$, then the induced metric $\eta_{ab}$ on the tangent space is a diagonal metric with signature $(4,3)$ and consequently $det(\eta)=-1$. The manifold $\mathcal{M}$ is a constant negative curvature symmetric manifold $H^{4,3}$ written as the coset $H^{4,3}=\f{SO(4,4)}{SO(4,3)}$  or a constant negative curvature non-symmetric coset space $H^{4,3}=\f{Spin(4,3)}{G_{2,2}}$ in the presence of torsion.
\end{itemize}
This is consistent with the discussion of parallelizability for both $S^7$, $H^{4,3}$ in the seminal papers by Cartan and Schouten \cite{Cartan-Schouten}, and Wolf \cite{Wolf1, Wolf2}.

As our last comment we point out that an application of this result as a way to construct Englert-type solutions of exotic $M'$-theory has been discussed in \cite{Henneaux:2017afd} where $S^{3,4}=\frac{Spin(3,4)}{G_{2,2}}$ appears as the internal manifold in dimensional reduction of exotic supergravity in $6+5$ dimensions (low energy limit of $M'$-theory).

Besides the potential implications for compactifications in string theory and supergravity, the fact that three-dimensional constant curvature manifolds are parallelizable may be of physical interest, like in the case described in \cite{Alvarez:2015bva}, where the Lorentz-flat condition in the 2+1 AdS geometry was exploited to analyze the coupling between the black hole and an $SU(2)$ gauge field. Also, in the standard big-bang cosmological model, in which the universe is conceived as a stack of three-dimensional constant curvature euclidean spaces labeled by cosmic time, the positive constant curvature spatial sections could also be described by a flat $SO(3)$ connection or by its torsion-free part.

%%%%%%%%%%%%%%%%%%%%%%%%%%%%%%%%%%
\subsection*{Acknowledgements}
%%%%%%%%%%%%%%%%%%%%%%%%%%%%%%%%%%

We are grateful to P.D.Alvarez, L.Castellani, B.L.Cerchiai, J.Edelstein, R.Emparan, M. Henneaux and K.Pilch for insightful discussions at different times in the development of this project. A.R. was partially supported by the ERC Advanced Grant `High-Spin-Grav', by F.R.S.-FNRS-Belgium (convention FRFC PDR T.1025.14 and  convention IISN 4.4503.15) and by the ``Communaut\'e Fran\c{c}aise de Belgique" through the ARC program. J.Z. has been partially funded by Fondecyt grant 1180368. The Centro de Estudios Cient\'ificos (CECs) is funded by the Chilean Government through the Centers of Excellence Base Financing Program of CONICYT-Chile. 

%%%%%%%%%%%%%%%%%%%%%%%%%%%%APPENDICES%%%%%%%%%%%%%%%%%%%%%%%%%

%%%%%%%%%%%%%%%%%%%%%%%%%%%%%%%%%%
\section*{Appendix. Division algebras and parallelizability of spheres}\label{Appendix:Division-algebra}
%%%%%%%%%%%%%%%%%%%%%%%%%%%%%%%%%%
\setcounter{equation}{0}  % reset counter
\renewcommand{\theequation}{A.\arabic{equation}}
As we mentioned the use of normed division algebras plays an important role providing us the sufficient condition for parallelizability of the manifold. For this reason it is helpful to have a quick review of what a division algebra is and how it can be related to the framing of a manifold.
An algebra $\mathfrak{g}$ is called a division algebra if for any non-zero element $a\in \mathfrak{g}$ there exist inverse elements for both left and right multiplications. If $\mathfrak{g}$  is also a normed vector space it is called a normed division algebra. There is a theorem by Hurwitz \cite{Hurwitz} which states that there are only four normed division algebras $\mathbb{R}$, $\mathbb{C}$, $\mathbb{H}$ and $\mathbb{O}$ where $\mathbb{R}$ and $\mathbb{C}$ are real and complex vector spaces respectively, $\mathbb{H}$ is the quaternion algebra and $\mathbb{O}$ is the octonion algebra. These division algebras have peculiar properties; $\mathbb{R}$ is ordered, commutative and associative algebra, $\mathbb{C}$ is a commutative and associative algebra, $\mathbb{H}$ is a non-commutative and associative algebra and $\mathbb{O}$ is a non-commutative and non-associative algebra. However, they are all alternative \cite{Schafer}, i.e. $\forall a,b \in \mathfrak{g}$
\begin{align}
a^2 b &= a(ab),\nn
b a^2 &= (ba)a,
\end{align}
and share a common and important property: they are \emph{normed} algebras. On the other hand, the possible dimension of any real division algebra is either $1$, $2$, $4$ or $8$ \cite{Hopf, Bott-Milnor, Kervaire}. Consequently, it is inferred that only normed division algebras of dimensions $1$, $2$, $4$ and $8$ are $\mathbb{R}$, $\mathbb{C}$, $\mathbb{H}$ and $\mathbb{O}$.

%%%%%%%%%%%%%%%%%%%%%%%%%%
\subsubsection*{Quaternions}
%%%%%%%%%%%%%%%%%%%%%%%%%%

A quaternion $q \in \mathbb{H}$ is determined by four real numbers $q^0, q^a$, $a=1,2,3 \;$, as $q=q^0 + q^a \lambda_a$, where the imaginary units $\lambda_a$ satisfy the rule given in Table \ref{Tab:Multipl-quat},
\be \label{q-lambdas}
\lambda_a \lambda_b = -\delta_{ab} + \epsilon_{abc} \lambda_c\,.
\ee
This multiplication rule (\ref{q-lambdas}) also defines the $SO(4)$-invariant bilinear norm for quaternions,
\be\label{|q2|}
|q^2|=q^* q= (q^0)^2+(q^1)^2+(q^2)^2+(q^3)^2,
\ee
where $q^*=q^0 - q^a \lambda_a$.

\begin{table}[htp]
\begin{center}
\begin{tabular}{|c|c|c|c|}
\hline \
       & $\lambda_1$ & $\lambda_2$ & $\lambda_3$ \\ 
\hline \
 $\lambda_1$ & $-1$  & $\lambda_3$ & $-\lambda_2$ \\
\hline  
 $\lambda_2$ & $-\lambda_3$ & $-1$ & $\lambda_1$ \\
\hline  
 $\lambda_3$ & $\lambda_2$ & $-\lambda_1$ & $-1$ \\
\hline
\end{tabular}
\caption{Multiplication table for imaginary quaternions.} \label{Tab:Multipl-quat}
\end{center}
\end{table}

The multiplication rule of imaginary quaternion units provides a rule for a cross product ($\times$) in $V^3$:
\be\label{x-product3}
(v\times w)_a = \epsilon_{abc} \, v_b \, w_c, \qquad \forall \,v,w, v\times w \in V^3 \,.
\ee 
Additionally, \eqref{|q2|} induces a positive definite $SO(3)$-invariant scalar product in $V^3$, $v\cdot w = v^a w_a$ given by the Euclidean metric $\delta_{ab}= diag(+,+,+)$.

%%%%%%%%%%%%%%%%%%%%%%%%%%
\subsubsection*{Split quaternions}
%%%%%%%%%%%%%%%%%%%%%%%%%%

Split quaternions, defined by four real numbers $\tilde{q}^0, \tilde{q}^a \in \mathbb{R}$ and three imaginary units $\tilde{\lambda}_a$, such that
\be\label{qtilde-lambdas}
\tilde{q}=\tilde{q}^0 + \tilde{q}^a \tilde{\lambda}_a\,,\quad \mbox{with} \quad \tilde{\lambda}_a \tilde{\lambda}_b = \eta_{ab} + \epsilon\indices{^c_{ab}} \tilde{\lambda}_c\,,
\ee
where $\eta_{ab}=diag(-1,+1,+1)$ and $\epsilon\indices{^a_{bc}} = \eta^{ad}\epsilon_{dbc}$ so that the imaginary units $\tilde{\lambda}_a$ satisfy the multiplication rule given in Table \ref{Tab:Multipl-squat}.
It turns out that there exists a cross product in three dimensional vector space $V^{2,1}$ denoted by $\times_s$ and defined as
\begin{equation}
(v \times_s  w)^a =\tilde{f}^a{}_{bc} v_b w^c,\qquad \, v, w, v \times_s w \in \textrm{Im}\,\mathbb{H}_s,
\end{equation}
which is now an invariant operation under $SO(2,1)$. We emphasize that $V^{2,1}\simeq \textrm{Im}\,\mathbb{H}_s$.

The $SO(2,1)$-invariant metric $\eta_{ab}$ is induced from the $SO(2,2)$-invariant bilinear form on the space of split-quaternions $\mathbb{H}_s$,
\be \label{|q-tilde|}
|\tilde{q}^2|=\tilde{q}^*\tilde{q}= (\tilde{q}^0)^2+(\tilde{q}^1)^2-(\tilde{q}^2)^2-(\tilde{q}^3)^2,
\ee
where $\tilde{q}^*=\tilde{q}^0-\tilde{q}^a \tilde{\lambda}_a$.

\begin{table}[htp]
\begin{center}
\begin{tabular}{|c|c|c|c|}
\hline \
   & $\tilde{\lambda}_1$ & $\tilde{\lambda}_2$ & $\tilde{\lambda}_3$ \\ 
\hline \
 $\tilde{\lambda}_1$ & $-1$  & $\tilde{\lambda}_3$ & $-\tilde{\lambda}_2$ \\
\hline  
 $\tilde{\lambda}_2$ & $-\tilde{\lambda}_3$ & $1$ & $-\tilde{\lambda}_1$ \\
\hline  
 $\tilde{\lambda}_3$ & $\tilde{\lambda}_2$ & $\tilde{\lambda}_1$ & $1$ \\
\hline
\end{tabular}
\caption{Multiplication table for imaginary \textit{split} quaternions.} \label{Tab:Multipl-squat}
\end{center}
\end{table}

%%%%%%%%%%%%%%%%%%%%%%%%%%
\subsubsection*{Octonions}
%%%%%%%%%%%%%%%%%%%%%%%%%%

The octonions are defined in a similar way \cite{Baez} with eight real numbers $z^0, z^a$, $a=1,\cdots,7 \;$, and seven imaginary units $\lambda^a$,  as $z=z^0 + z^a \lambda_a$, where the imaginary units satisfy the multiplication rule given in Table \ref{Tab:Multipl-Oct},
\be \label{o-lambdas}
\lambda_a \lambda_b = -\delta_{ab} + f_{abc} \lambda_c\,,
\ee
where $f_{abc}$ is totally antisymmetric in its indices and
\begin{equation} \label{f}
f_{abc}=+1\,, \mbox{for    } abc=123, 145, 176, 246, 257, 347, 365.
\end{equation}

\begin{figure}[htp]
\begin{center}
\def\svgscale{0.6}
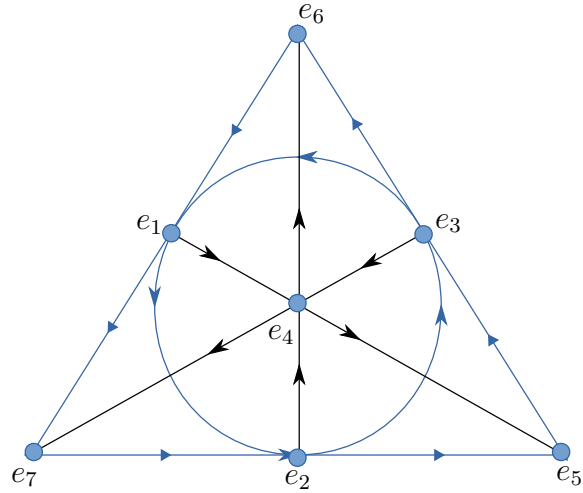
\caption{The Fano plane for octonions which is a mnemonic way of finding totally antisymmetric $f_{abc}$ instead of using the multiplication table \ref{Tab:Multipl-Oct}.}\label{Fig:Fano_oct}
\end{center}
\end{figure}

The coefficients $f_{abc}$ are defined by the multiplication table \ref{Tab:Multipl-Oct} for imaginary octonions.\footnote{These coefficients can be read off from the Fano plane in Figure \ref{Fig:Fano_oct}. Note that this is only one choice of octonion multiplication table.}

\begin{table}[htp]
\begin{center}
\begin{tabular}{|c|c|c|c|c|c|c|c|}
\hline \
       & $\lambda_1$ & $\lambda_2$ & $\lambda_3$ & $\lambda_4$ & $\lambda_5$ & $\lambda_6$ & $\lambda_7$  \\ 
\hline \
 $\lambda_1$ & $-1$  & $\lambda_3$ & $-\lambda_2$ & $\lambda_5$ & $-\lambda_4$ & $-\lambda_7$ & $\lambda_6$  \\
\hline  
 $\lambda_2$ & $-\lambda_3$ & $-1$ & $\lambda_1$ & $\lambda_6$ & $\lambda_7$ & $-\lambda_4$ & $\lambda_5$  \\
\hline  
 $\lambda_3$ & $\lambda_2$ & $-\lambda_1$ & $-1$ & $\lambda_7$ & $-\lambda_6$ & $\lambda_5$ & $-\lambda_4$  \\
\hline  
 $\lambda_4$ & $-\lambda_5$ & $-\lambda_6$ & $-\lambda_7$ & $-1$ & $\lambda_1$ & $\lambda_2$ & $\lambda_3$  \\
\hline  
 $\lambda_5$ & $\lambda_4$ & $-\lambda_7$ & $\lambda_6$ & $-\lambda_1$ & $-1$ & $-\lambda_3$ & $\lambda_2$  \\
\hline  
 $\lambda_6$ & $\lambda_7$ & $\lambda_4$ & $-\lambda_5$ & $-\lambda_2$ & $\lambda_3$ & $-1$ & $-\lambda_1$  \\
\hline  
 $\lambda_7$ & $-\lambda_6$ & $\lambda_5$ & $\lambda_4$ & $-\lambda_3$ & $-\lambda_2$ & $\lambda_1$ & $-1$  \\
\hline
\end{tabular}
\caption{Multiplication table for imaginary octonions ($\textrm{Im}\,\mathbb{O}$). The elements $\lambda_1,...,\lambda_7$ form a basis for $\textrm{Im}\,\mathbb{O}$.} \label{Tab:Multipl-Oct}
\end{center}
\end{table}

The cross product $\times$ on the seven dimensional vector space $V^7$, isomorphic to the space of imaginary octonions $\textrm{Im}\, \mathbb{O}$, is defined as
\be\label{x-product7}
(v \times w)^a =f^a{}_{bc} v^b w^c,       
\ee
where $v, w, v \times w \in \textrm{Im}\, \mathbb{O}$.

%%%%%%%%%%%%%%%%%%%%%%%%%%
\subsubsection*{Split octonions}
%%%%%%%%%%%%%%%%%%%%%%%%%%

\begin{figure}[htp]
\begin{center}
\def\svgscale{0.6}
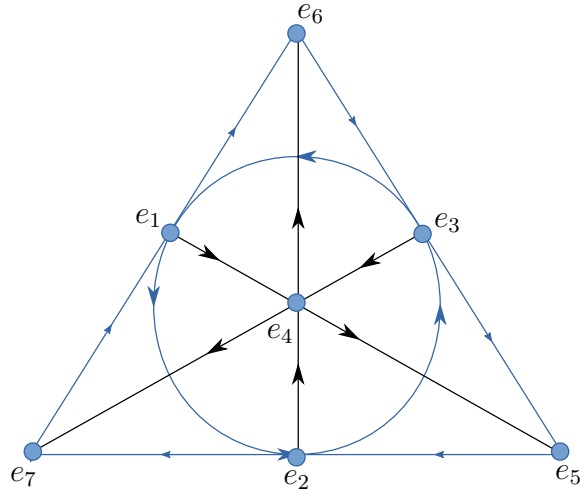
\caption{The Fano plane for split-octonions which is a mnemonic way of finding totally antisymmetric $\tilde{f}_{abc}$ instead of using the multiplication table \ref{Tab:Multipl-Split-Oct}.}\label{Fig:Fano_split}
\end{center}
\end{figure}

The split octonions are also defined by eight real numbers $\tilde{z}^0, \tilde{z}^a$, $a=1,\cdots,7 \;$, and seven imaginary units $\tilde{\lambda}^a$,  as $\tilde{z}=\tilde{z}^0 + \tilde{z}^a \tilde{\lambda}_a$. The imaginary units satisfy
\begin{equation}\label{split-product}
\lambda_a \lambda_b = \eta_{ab} + \tilde{f}^c{}_{ab} \lambda_c \,
\end{equation}
where $SO(4,3)$ invariant metric $\eta_{ab}= diag(-,-,-,+,+,+,+)$ and its inverse are used to lower and raise the indices, and $\tilde{f}_{abc} \equiv \eta_{ad} \tilde{f}^d{}_{bc}$ is totally antisymmetric, taking the values $\pm 1, 0$, with
\begin{equation*}
    f_{abc}=+1, \mbox{   for   } abc=123,145,167,246,275,347,356.
\end{equation*}
Formally we have
\be
f_{abc}=-\f{1}{3} \left(\eta_{ad} f\indices{^d_{bc}} + \eta_{bd} f\indices{^d_{ca}} + \eta_{cd} f\indices{^d_{ab}} \right).
\ee
This is summarized in Table \ref{Tab:Multipl-Split-Oct} of imaginary split-octonions, and the corresponding Fano plane in Figure \ref{Fig:Fano_split}.

\begin{table}[htp]
\begin{center}
\begin{tabular}{|c|c|c|c|c|c|c|c|c|}
\hline 
      & $\lambda_1$ & $\lambda_2$ & $\lambda_3$ & $\lambda_4$ & $\lambda_5$ & $\lambda_6$ & $\lambda_7$ \\
\hline  
$\lambda_1$ & $-1$ & $\lambda_3$ & $-\lambda_2$ & $-\lambda_5$ & $\lambda_4$ & $-\lambda_7$ & $\lambda_6$ \\ 
\hline  
$\lambda_2$ & $-\lambda_3$ & $-1$ & $\lambda_1$ & $-\lambda_6$ & $\lambda_7$ & $\lambda_4$ & $-\lambda_5$ \\
\hline  
$\lambda_3$ & $\lambda_2$ & $-\lambda_1$ & $-1$ & $-\lambda_7$ & $-\lambda_6$ & $\lambda_5$ & $\lambda_4$ \\
\hline  
$\lambda_4$ & $\lambda_5$ & $\lambda_6$ & $\lambda_7$ & $1$ & $\lambda_1$ & $\lambda_2$ & $\lambda_3$ \\
\hline  
$\lambda_5$ & $-\lambda_4$ & $-\lambda_7$ & $\lambda_6$ & $-\lambda_1$ & $1$ & $\lambda_3$ & $-\lambda_2$ \\
\hline  
$\lambda_6$ & $\lambda_7$ & $-\lambda_4$ & $-\lambda_5$ & $-\lambda_2$ & $-\lambda_3$ & $1$ & $\lambda_1$ \\
\hline  
$\lambda_7$ & $-\lambda_6$ & $\lambda_5$ & $-\lambda_4$ & $-\lambda_3$ & $\lambda_2$ & $-\lambda_1$ & $1$ \\
\hline
\end{tabular}
\caption{The multiplication table for imaginary split-octonion algebra $\textrm{Im}\,\mathbb{O}_s$. The elements $\lambda_1,...,\lambda_7$ form a basis for  $\textrm{Im}\,\mathbb{O}_s$.}\label{Tab:Multipl-Split-Oct}
\end{center}
\end{table}

%%%%%%%%%%%%%%%%%%%%%%%%%%

\subsubsection*{Relation to parallelizability}

One way to relate the division algebras to parallelizability of spheres is through the projective geometry. It was shown that for any normed division algebra $\mathbb{G}$, the projective line $\mathbb{GP}^1$ is isomorphic to a sphere of the dimension of $\mathbb{G}$, e.g. $\mathbb{OP}^1 \cong S^8$. Any projective line $\mathbb{GP}^1$ comes along with a canonical $n$ dimensional line bundle $L_{\mathbb{G}}$ or equivalently a unit ($n-1$)-sphere bundle (Hopf bundle) at each point on the $n$ dimensional base manifold $\mathbb{GP}^1$. This allows one to introduce a Hopf map from a $2n-1$ dimensional total manifold to $\mathbb{GP}^1$ which includes an ($n-1$)-sphere bundle at each point. Using the Hopf invariant constructed from Hopf maps, it has been shown that for an $n$ dimensional normed division algebra, $S^{n-1}$ is parallelizable \cite{Hopf, Adams}.

The tangent space of any unit ($n-1$)-sphere can be thought of as an $n-1$ dimensional vector space $\textrm{Im}\,\mathbb{G}$ constructed from imaginary elements of corresponding normed division algebra $\mathbb{G}$. As a consequence one can define a Lie algebra on the tangent space with a Lie group realized as the group $\textrm{Aut}(\mathbb{G})$ of (outer) automorphism of $\mathbb{G}$. For $\mathbb{R}$ and $\mathbb{C}$ the automorphism groups are trivial. For quaternions, $\textrm{Aut}(\mathbb{H})\cong SO(3)$ and for octonions $\textrm{Aut}(\mathbb{O})\cong G_2$.
%\end{appendix}

%%%%%%%%%%%%Bibliography%%%%%%%%%%%%%

\end{document}